\documentclass[twocolumn,superscriptaddress,prl,amsmath,amssymb]{revtex4-2}

\usepackage{graphicx}
\usepackage{color}
\usepackage{amsmath}
\usepackage{bm}
\usepackage{ulem}
\usepackage{hyperref}

\begin{document}

\title{Localization Landscape in Non-Hermitian and Floquet quantum systems}


\author{David Guéry-Odelin}
\affiliation{Laboratoire Collisions Agrégats Réactivité, UMR 5589, FERMI, Université de Toulouse, CNRS, 118 Route de Narbonne, 31062 Toulouse CEDEX 09, France.}
`
\author{François Impens}
\affiliation{Instituto de F\'{\i}sica, Universidade Federal 
Rio de Janeiro, 21941-972 Rio de Janeiro, RJ, Brazil}

\date{\today}

\begin{abstract}
We propose a generalization of the Filoche--Mayboroda localization landscape that extends the theory well beyond the static, elliptic and Hermitian settings while preserving its geometric interpretability. Using the positive operator $H^\dagger H$, we obtain a landscape that predicts localization across non-Hermitian, Floquet, and topological systems without computing eigenstates. Singular-value collapse reveals spectral instabilities and skin effects, the Sambe formulation captures coherent destruction of tunneling, and topological zero modes emerge directly from the landscape. Applications to Hatano--Nelson chains, driven two-level systems, and driven Aubry--André--Harper models confirm quantitative accuracy, establishing a unified predictor for localization in equilibrium and driven quantum matter.
\end{abstract}

\maketitle 
Localization phenomena lie at the heart of quantum mechanics and wave physics, governing the spatial confinement of eigenstates in disordered, quasiperiodic, or topologically nontrivial systems~\cite{Anderson1958,Anderson79,AubryAndre1980,Roati2008,Schreiber2015,White2020Anderson2D}. Understanding where and why eigenfunctions concentrate has profound implications for transport, thermalization, and the emergence of insulating behavior in condensed matter~\cite{Basko2006,Nandkishore2015}, as well as for the design of photonic and phononic metamaterials. Beyond equilibrium settings, driven quantum systems exhibit rich localization physics, including dynamical localization in periodically modulated lattices~\cite{Eckardt05,Lignier2007,Eckardt2007,floquet_review,Goldman2014,Holthaus2016,LocalizationZhangNJP24,Ponte2015,Lazarides2015,Bordia2017,Crowley2020}.
Similarly, non-Hermitian quantum systems exhibit exotic boundary phenomena such as
the non-Hermitian skin effect~\cite{SKin1,Skin2,Skin3,Skin4,SkinLonghi1,SkinReview,SkinPRX25}
and display counterintuitive topological features~\cite{EPReview,Li2023ExceptionalPoints}. 
Transport in non-Hermitian systems~\cite{hatano1996,Eichelkraut2013Mobility,NHTLonghi17,Xu2023NonHermitianMetagratings} and topological materials~\cite{RMPToplogicalBand} offer exciting perspectives.
This context calls for a unified geometric framework capable of capturing the
interplay between localization and disorder in non-Hermitian, Floquet and topological quantum systems.

In its current form, the localization landscape theory introduced by Filoche and Mayboroda~\cite{filoche2012} and its subsequent developments~\cite{Arnold2016,LandscapeManyBody,Sergey24} offer a powerful geometric framework. 
The localization landscape $u$, defined as the unique solution of $H u = 1$ under
Dirichlet boundary conditions, encodes confinement properties without requiring the
explicit diagonalization of $H$. Its reciprocal $1/u$ acts as an effective
confining potential~\cite{Arnold2016}, offering a quantitative description of
strong localization~\cite{Arnold2016,Gontier2019} in terms of effective quantum
tunneling and providing rigorous bounds on eigenfunction confinement.
This framework has proven remarkably versatile, extending to Anderson localization~\cite{filoche2012}, wave confinement in complex media~\cite{lefebvre2016}, semiconductors~\cite{LandscapeSC1,LandscapeSC2,LandscapeSC3}, Bose Einstein condensates~\cite{LandscapeBEC1,LandscapeBEC2} and metamaterial design~\cite{dalnegro2024}.
A key practical advantage of the landscape approach is its predictive power and computational efficiency: while 
full diagonalization scales as $O(N^3)$, the landscape equation 
can be solved using sparse iterative methods with quasi-linear 
scaling. Despite its wide applicability, localization landscape theory has remained fundamentally restricted to static, elliptic operators whose positivity and self-adjoint structure guarantee a unique positive landscape, thereby supporting its interpretation as an effective confining potential. These mathematical properties 
are naturally satisfied in stationary Schrödinger operators and classical wave equations, 
but they are generally lost in non-Hermitian or explicitly time-dependent systems.

In this Letter, we generalize the localization landscape framework to 
non-Hermitian and Floquet systems by defining $v(x)$ as the solution of the equation
\begin{equation}
H^\dagger H\, v = \mathbf{1}
\label{eq:gen_landscape}
\end{equation}
The operator $H^\dagger H$ preserves the essential features---positivity and Hermiticity---needed to infer localization, even when $H$ itself lacks them in the physical contexts discussed below. At the same time, its eigenstates remain closely related to those of $H$ and even coincide when $H^\dagger = H$. 
When $H$ is Hermitian and positive definite, Eq.~\eqref{eq:gen_landscape} reduces to $H^2 v = \mathbf{1}$, so that $v = H^{-2}\mathbf{1} = H^{-1}u$, with $u$ the conventional landscape defined by $Hu = \mathbf{1}$. 
The generalized landscape is therefore a smoothed version of the standard one, fully compatible with the usual elliptic framework. It also preserves the key inequality governing localization: any eigenmode $\varphi(x)$ of $H^\dagger H$ with eigenvalue $\lambda = E^2 \ge 0$ satisfies $|\varphi(x)| \le E^2 \|\varphi\|_\infty\, v(x)$,
which confines the eigenmodes to the regions bounded by the valley lines of $v(x)$. This construction also offers a natural extension of the localization landscape to non-Hermitian, or even non-elliptic, Hamiltonians.

A central advantage of this generalized landscape is its intrinsic sensitivity to spectral gap closings. 
When a control parameter approaches a value for which the operator $H$ develops a near-zero singular value, the smallest singular value $\sigma_{\min}(H)$ tends to zero.
Since the landscape equation involves the inverse of $H^\dagger H$, whose smallest eigenvalue is $\sigma_{\min}(H)^2$, the landscape amplitude necessarily becomes large.
More precisely, one has the bound
$v_{\max} \le \|v\|_2 \le \sqrt{d}\,\sigma_{\min}(H)^{-2}$,
where $d$ denotes the dimension of the truncated Hilbert space.
As a result, the generalized landscape provides a direct geometric diagnostic of resonant regimes, spectral instabilities and localization phenomena associated with near-kernel modes of $H$. An other major benefit of our approach is to address localization not only in static, but also in time-periodic (Floquet) quantum systems as explained below. 

Consider a non-Hermitian system, for which the positive Hermitian operator $H^\dagger H$ remains well defined and provides a natural starting point for extending the landscape construction.
To handle non-invertible or defective operators in a controlled manner, we invoke the Moore--Penrose pseudoinverse of $H^\dagger H$, which preserves both positivity and the geometric interpretability of the landscape.
Within this framework, we show that the generalized landscape naturally captures boundary accumulation phenomena and the non-Hermitian skin effect. 

\begin{figure}[t]
    \centering
    \includegraphics[width=0.98\columnwidth]{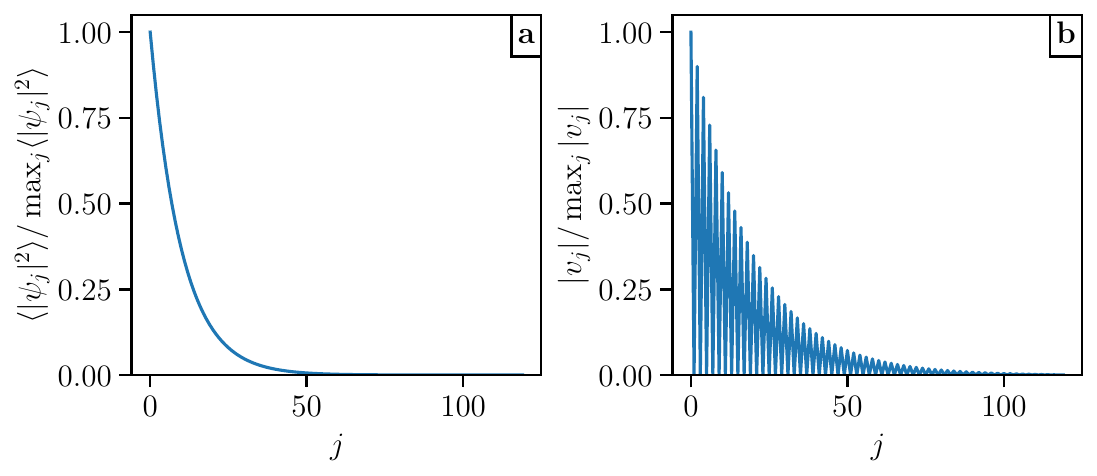}
    \caption{
Generalized landscape for the Hatano--Nelson chain under open boundary 
conditions, shown here for the regime $r=t_R/t_L<1$. 
Panel (a): normalized average right-eigenstate density profile $\langle |\psi_j|^2\rangle /\max_j \langle |\psi_j|^2\rangle$ (see text for definition). 
Panel (b): normalized landscape profile $|v_j|/\max_j |v_j|$. 
Parameters: $N=120$, $t_L=1$, $r=0.9$.
}
    \label{fig:hn_v_profiles}
\end{figure}

As a benchmark, we consider the
Hatano-Nelson model~\cite{hatano1996} with non-reciprocal nearest-neighbor
hopping, $H_{\rm HN}=\sum_{j=1}^{N-1}\Big(t_R\,|j{+}1\rangle\langle j|
+ t_L\,|j\rangle\langle j{+}1|\Big)$.
Under open boundary conditions, the hopping asymmetry $r=t_R/t_L$ induces a
macroscopic accumulation of eigenstates at one boundary~\cite{SkinReview}, realizing the
non-Hermitian skin effect: for $r<1$ ($r>1$), all right eigenstates localize
exponentially near the left (right) edge of the chain. In Fig.~\ref{fig:hn_v_profiles}a we show the average density
\[
\langle |\psi_j|^2\rangle = \frac{1}{N}\sum_{k=1}^N
\frac{|\psi^{(k)}_j|^2}{\sum_{\ell=1}^N |\psi^{(k)}_\ell|^2},
\]
obtained by averaging over all normalized right eigenstates, which clearly illustrates this behavior. To capture this localization geometrically, we construct the generalized
landscape $v = (H_{\rm HN}^\dagger H_{\rm HN})^{+}\mathbf{1}$, where
$(\cdot)^{+}$ denotes the Moore--Penrose pseudoinverse~\cite{FootnoteLandscape1}.
The spatial dependence of $|v_j|$ defines an effective geometric field along the
chain. The behavior of this landscape for the regime $r<1$ is shown in
Fig.~\ref{fig:hn_v_profiles}b. Both quantities exhibit a pronounced peak at the left boundary,
matching the expected direction of the non-Hermitian skin effect.
By symmetry, for $r>1$, the same localization patterns
appear, but mirrored with respect to the chain, with peaks shifted to the right
boundary. A quantitative correlation analysis 
shows that
the soft geometric indicators extracted from the generalized landscape provide a
remarkably accurate predictor of the collective displacement of the spectrum. In particular, the soft centers of mass $x_v^{(1)}$, defined from weighted averages of $|v_j|$, exhibit consistently strong agreement with the eigenstate center of mass $\overline{x}_{\rm cm}$. Throughout the range of hopping ratios $r \in [0.7,1.3]$, both quantities show Pearson correlation coefficients of $\approx 0.99$ and Spearman coefficients essentially equal to $1.0$, demonstrating a robust and strictly monotonic relationship even across the strongly non-Hermitian regime.
The soft landscape predictors remain
reliable even in the vicinity of the crossover where the landscape becomes nearly
flat. The center of mass of the generalized landscape therefore
captures the correct boundary-localization geometry of the non-Hermitian skin.%

This non-Hermitian example shows that the $H^\dagger H$ formulation preserves the geometric content of the localization landscape while extending it to operators that are non-normal or singular under open boundary conditions. Having established its relevance in static non-Hermitian settings, we now turn to a second class of systems that exhibits dynamical localization. A paradigmatic example is coherent destruction of tunneling (CDT)~\cite{Grossmann91,HanggiPRL91}, where a strongly driven system becomes effectively frozen despite large time-periodic modulation, illustrating how transport suppression can emerge from interference in time rather than from a static potential landscape. 

Such dynamical localization mechanisms can be recast within a non-Hermitian framework via Floquet theory~\cite{sambe1973}. For a time-periodic Hamiltonian $H(t)= H(t+T)$ with period $T=2\pi/\omega$, Floquet's theorem ensures the existence of solutions of the form $|\psi(t)\rangle = e^{-i\varepsilon t/\hbar}|u_\varepsilon(t)\rangle$, where the Floquet modes $|u_\varepsilon(t)\rangle$ are $T$-periodic. Expanding these modes in Fourier harmonics, $|u_\varepsilon(t)\rangle=\sum_{m=-\infty}^{\infty}e^{im\omega t}|u_\varepsilon^{(m)}\rangle$, maps the problem onto a stationary eigenvalue equation in the extended Sambe space $\mathcal H_S=\mathcal H\otimes\mathcal H_{\mathrm{Fourier}}$. The resulting Sambe operator $H_S=H(t)-i\hbar\partial_t$ is by construction non-Hermitian, placing driven systems on the same mathematical footing as the static non-Hermitian problems discussed previously. Its matrix elements read
\[
\langle j,m|H_S|j',m'\rangle = \langle j|H_{m-m'}|j'\rangle + m\hbar\omega\,\delta_{jj'}\delta_{mm'},
\]
where off-diagonal blocks couple different harmonic sectors, while the diagonal term $m\hbar\omega$ generates a ladder of quasi-energy replicas. In practice, the Fourier expansion is truncated to $m\in[-M,M]$, yielding a finite-dimensional matrix of dimension $d=N(2M+1)$.

Temporal confinement arises naturally within the extended space, connecting Floquet phenomena~\cite{floquet_review,Rahav2003} to landscape geometry and to effective Hamiltonian methods~\cite{HanggiPRL91,Breuer1991}. The framework proves particularly effective for analyzing heating~\cite{Lazarides2014,DAlessio2014} and multi-frequency protocols~\cite{Ikeda2020}. We illustrate these capabilities with two representative examples: driven two-level systems with multiple frequencies and the driven Aubry–André–Harper chain~\cite{Aubry1980,Luo2005}. These applications show how the generalized landscape reveals dynamical and temporal localization regimes~\cite{Haldar2021}, establishing it as a versatile diagnostic for Floquet engineering~\cite{Holthaus2016,Rudner2020} and periodically modulated quantum matter~\cite{Goldman2014,Mikami2016}.

Within this static Sambe representation, quasi-energy gap closings acquire a transparent
algebraic meaning. If, for a control parameter value $A=A_\star$, a quasi-energy
$\varepsilon_\star(A)$ approaches zero, there exists a normalized vector
$\phi_\star$ such that
$H_S(A_\star)\phi_\star=\varepsilon_\star(A_\star)\phi_\star$ with
$|\varepsilon_\star(A_\star)|\ll\omega$.
By the variational definition of the smallest singular value,
$\sigma_{\min}(H_S)=\min_{\|x\|_2=1}\|H_Sx\|_2$,
this immediately implies
$\sigma_{\min}(H_S(A_\star))\le|\varepsilon_\star(A_\star)|$.
Hence, a quasi-energy gap closing at $\varepsilon=0$ necessarily corresponds to the
emergence of an almost-null direction in Sambe space and to an ill-conditioned
Sambe operator. The dynamical significance follows directly.
The stroboscopic evolution over one period is governed by
$U(T)=e^{-iH_FT}$, and for $|\varepsilon_\star|\ll\omega$ the phase factor
$e^{-i\varepsilon_\star T}$ is close to unity.
Floquet modes associated with small quasi-energies therefore evolve anomalously
slowly at stroboscopic times, leading to dynamical freezing and localization.
In Sambe space, this slowdown is precisely encoded in the singular structure of
$H_S$, independently of detailed spectral properties.

Rather than working explicitly with Floquet eigenstates, we exploit the singular structure of the Sambe operator by focusing on the positive Hermitian operator $H_S^\dagger H_S$ and defining the Sambe localization landscape (SLL) as the solution of
$H_S^\dagger H_S\,v=\mathbf{1}$.
Within this formulation, quasi-energy gap closings manifest themselves geometrically through the amplification of the landscape response along near-null directions of $H_S$, providing a direct signature of resonant regimes and slow dynamics.

\begin{figure}[t]
    \centering
    \includegraphics[width=\columnwidth]{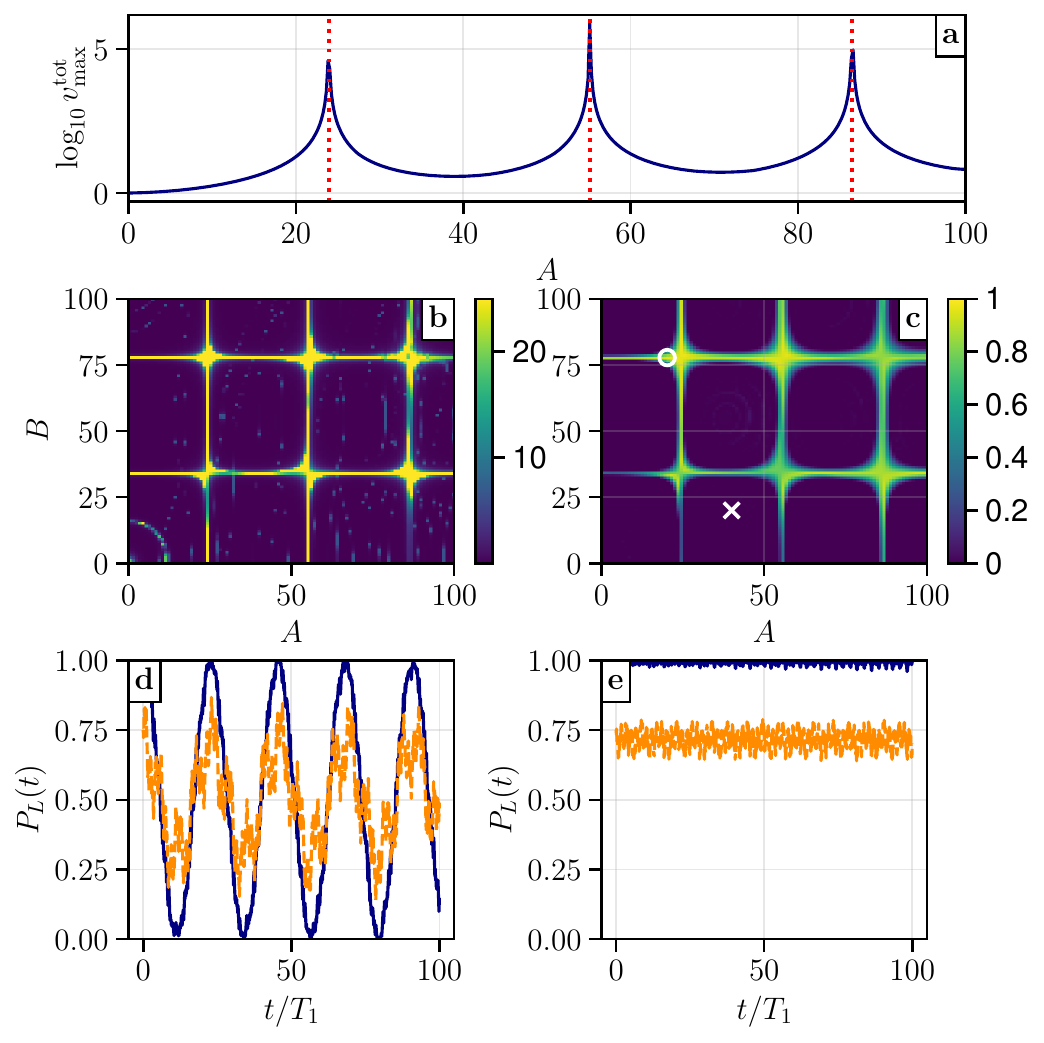}
    \caption{Sambe localization landscape for a driven two-level system. 
(a) Maximal landscape amplitude $v_{\max}^{\mathrm{tot}}(A)$ versus drive amplitude $A$ for monochromatic driving at frequency $\Omega$. Vertical dashed lines: CDT values obtained numerically from Floquet theory. 
(b) Two-dimensional landscape $v_{\max}^{\rm tot}(A,B)$ for a bichromatic drive with 
incommensurate frequencies $(\Omega_1, \Omega_2=\sqrt{2}\Omega_1)$ as a function 
of modulation amplitudes $A$ and $B$. The color scale shows $v_{\max}/1000$ for 
readability.
(c) Minimum left-site population $\min_t P_L(t;A,B)$ from direct time integration over 100 periods, starting from the left-localized state. 
(d),(e): Time evolution of the left-site population $P_L(t)$ for two different initial 
states: a fully left-localized state (blue), and a partially left-localized 
state $|\psi(0)\rangle = (\sqrt{3}\,|L\rangle + |R\rangle)/2$ (orange). 
The two figures (d) and (e) correspond to two distinct parameter sets $(A,B)$ indicated by 
the white cross and white circle in (c), illustrating respectively a 
delocalized dynamical regime (d) and a time-localized regime (e).
}
    \label{fig:CDTrecap}
\end{figure}
We first apply the SLL approach to a driven two-level system $\hat{H}(t) = -J\sigma_x + s(t)\sigma_z/2$. For monochromatic driving $s(t) = A\cos(\Omega t)$, the system exhibits CDT for a discrete set of values of the ratio $A/\hbar\Omega$~\cite{HanggiPRL91,Grossmann91,Lignier2007}. In the high-frequency limit, this physical effect is captured by an effective Hamiltonian $\hat{H}_{\mathrm{eff}} = -J_{\mathrm{eff}}\sigma_x$ with $J_{\mathrm{eff}} = J\,J_0(A/\hbar\Omega)$~\cite{Eckardt05}, where $J_0$ is the Bessel function of the first kind. CDT occurs when $J_0(A/\hbar\Omega) = 0$. 
Computing the SLL and tracking the quantity $v_{\max}^{\text{tot}}(A) = \max_j |v_j|$,
yields a sequence of sharp, well-resolved peaks (Fig.~\ref{fig:CDTrecap}a) that 
coincide with the exact CDT values extracted from the Floquet quasienergy gap.
This agreement persists across the full parameter range explored, demonstrating that the landscape provides a nonperturbative and geometrically transparent indicator of coherent tunnelling suppression. 
The landscape construction converges extremely rapidly: while the third peak shows an $8.7\%$ offset for $M=4$, the results for $M\ge 6$ become numerically indistinguishable from exact Floquet theory at machine precision. These observations indicate that only a small number of harmonics is required to fully capture the CDT structure.


For bichromatic driving $s(t) = A\cos(\Omega_1 t) + B\cos(\Omega_2 t)$ with incommensurate frequencies $\Omega_2/\Omega_1 = \sqrt{2}$, standard Floquet theory no longer applies. Nevertheless, the generalized Sambe landscape extends naturally to this quasiperiodic regime by working in an extended space $\mathcal{H} \otimes \ell^2(\mathbb{Z}^2)$ with two harmonic indices. The resulting two-dimensional map $v_{\max}^{\mathrm{tot}}(A,B)$ [Fig.~\ref{fig:CDTrecap}(b)] displays well-separated regions of enhanced and suppressed landscape intensity. 
These regions accurately predict dynamical localization, as confirmed by direct comparison with the minimum population $\min_t P_L(t)$ obtained from full time integration starting from an initial state localized on the left site [Fig.~\ref{fig:CDTrecap}(c)].
Across the entire $(A,B)$ parameter plane, correlations between the landscape
$v^{\rm tot}_{\max}(A,B)$ and the minimal left-site population $\min_t P_L(t;A,B)$ remain
remarkably strong, with Pearson coefficients close to $0.90$ and Spearman
coefficients around $0.58$. As expected, Figs.~\ref{fig:CDTrecap}(d,e) show that
these predictive capabilities are robust with respect to the choice of
initial state (fully localized or partially delocalized): in smooth regions
the left population exhibits coherent oscillations, whereas near landscape
maxima it remains strongly confined. This high level of concordance confirms
the reliability and generality of the landscape approach.

\begin{figure}[t]
    \centering
    \includegraphics[width=\columnwidth]{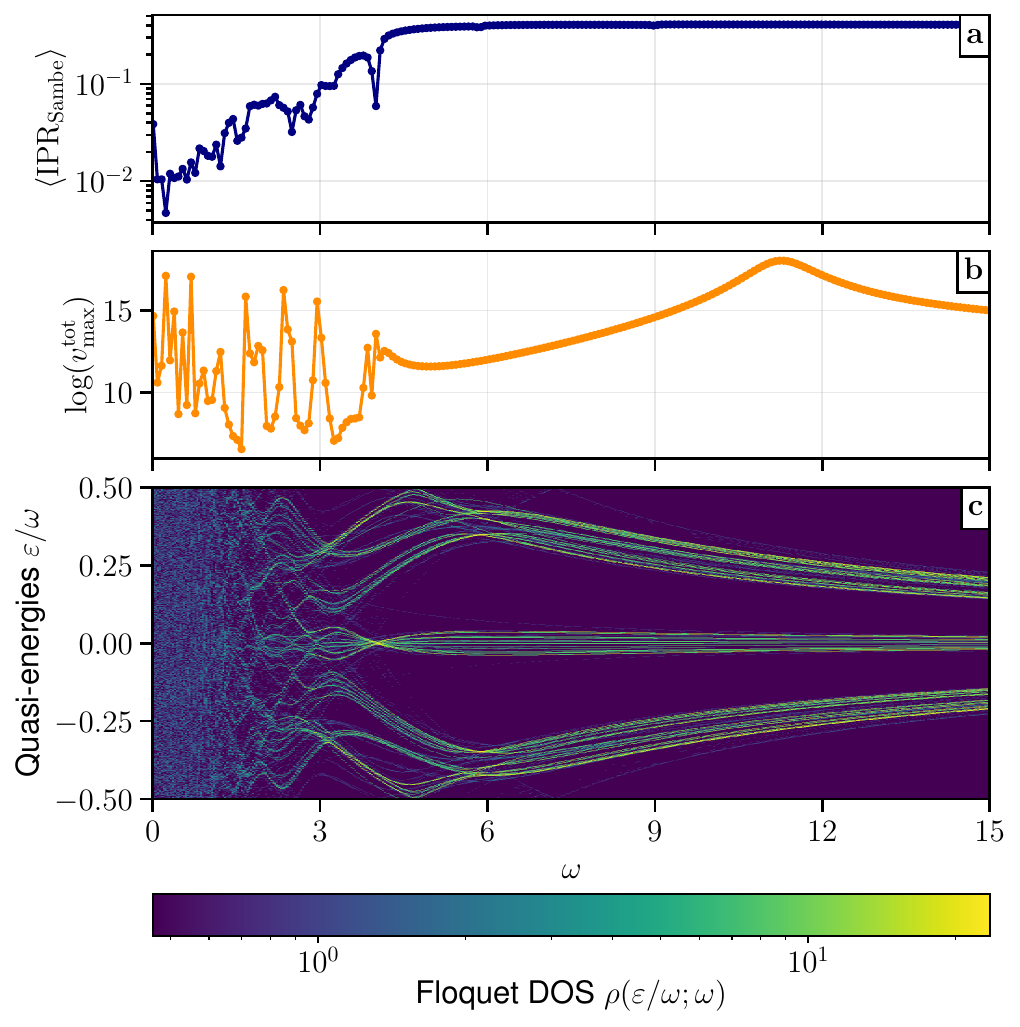}
    \caption{Floquet localization and spectral diagnostics for the driven Aubry--André--Harper chain at fixed amplitude $A \simeq 3.7$. (a) Sambe-space inverse participation ratio versus driving frequency $\omega$, showing irregular fluctuations at low $\omega$ and smooth high-frequency behavior. (b) Maximal landscape amplitude $v_{\max}^{\mathrm{tot}}(\omega)$, revealing near-singularities at low $\omega$. (c) Floquet density of states versus $\omega$ and scaled quasi-energy $\varepsilon/\omega$, correlating spectral rearrangements with localization diagnostics. Parameters: $N=80$, $J=1$, $\alpha=(\sqrt{5}-1)/2$ and $\lambda_0=2.8$.}
    \label{fig:cut_ipr_landscape}
\end{figure}

We next consider a driven Aubry--André--Harper chain with time-periodic modulation of the onsite potential: $H(t) = -J\sum_n (|n+1\rangle\langle n| + \mathrm{h.c.})+ [\lambda_0 + A\cos(\omega t)]\sum_n \cos(2\pi\alpha n + \theta)|n\rangle\langle n|$,
where $\alpha$ is irrational and $\lambda_0 > 2J$ places the system in the localized regime. The Sambe-space inverse participation ratio
$\mathrm{IPR}_{\mathrm{Sambe}} = \sum_{m,n} |\Psi(m,n)|^4$
probes localization jointly in real space and harmonic index.
At fixed drive amplitude $A\simeq 3.7$, the IPR as a function of the driving 
frequency $\omega$ [Fig.~\ref{fig:cut_ipr_landscape}(a)] exhibits pronounced, 
irregular fluctuations at low frequencies $\omega \leq 4$ due to 
resonant hybridization of Floquet replicas, and becomes 
smooth in the high-frequency regime where replicas decouple. 
The maximal landscape amplitude $v_{\max}^{\mathrm{tot}}(\omega)$ 
[Fig.~\ref{fig:cut_ipr_landscape}(b)] shows corresponding peaks at low $\omega$, 
signaling near-singularities of $H_S$ associated with quasi-energy gap closings. 
The underlying spectral reorganization is captured by the Floquet density of states 
$\rho(x;\omega)$ [Fig.~\ref{fig:cut_ipr_landscape}(c)], obtained by folding the 
Sambe eigenvalues $E_n(\omega)$ into $\varepsilon_n\in[-\omega/2,\omega/2)$, 
rescaling $x_n=\varepsilon_n/\omega$, and estimating $\rho(x;\omega)$ via a 
normalized histogram of $\{x_n\}$ with bin width $\Delta x$ 
(i.e., $\Delta\varepsilon=\omega\,\Delta x$). 
The DOS displays strong redistributions at low $\omega$ and stabilizes at high frequency. These results demonstrate that the SLL simultaneously characterizes Floquet-modified localization and provides a sensitive probe of quasi-energy structure.

In topological models, the non-positivity of $H$ invalidates the original landscape construction. In contrast, our new framework successfully recovers the ability to detect topological zero modes and determine their spatial location. Whenever a chiral-symmetry-protected near-zero mode is present, the smallest
eigenvalues of $H^\dagger H$ collapse and the landscape develops a sharp peak
precisely at the position where the topological state localizes—either at a
boundary or at an interface—highlighting the strong correlation between LDOS
and the localization landscape.
In the SSH chain~\cite{SSH79}, we verified this behavior across three distinct real-space configurations—topological, trivial, and a trivial–topological domain wall—whose low-energy spectra respectively host two midgap boundary modes, none, and a single midgap interface mode. Likewise, in the BBH higher-order topological phase~\cite{BBH1,BBH2}, the $H^\dagger H$ landscape
faithfully reproduces the expected corner localization, developing sharp peaks at the positions of the four midgap corner states.

In conclusion, the generalized $H^\dagger H$ landscape offers a unified geometric framework for diagnosing localization across a broad class of quantum systems, including regimes where standard methods fail. It naturally incorporates non-Hermitian physics—including the skin effect—and extends to Floquet systems by accounting for the synthetic Sambe dimensions associated with the driving frequencies, thus capturing both static and dynamical localization. Its predictive power is validated by excellent quantitative agreement with direct numerical simulations on paradigmatic models (CDT in driven two-level systems, Hatano–Nelson, AAH, and SSH chains).

Beyond efficient spectral diagnostics for large-scale simulations and Floquet or topological engineering, the approach also suggests a promising direction: inverse landscape engineering, in which disorder is deliberately shaped to tune the $H^\dagger H$ landscape—flattening gradients to enhance transport or steepening barriers to enforce localization. This perspective elevates the landscape from a diagnostic or predictive tool to a potential control parameter, opening design strategies for equilibrium and nonequilibrium quantum matter.

\paragraph{Acknowledgements.} This work was supported by the Institut Universitaire de France. F.I. acknowledges support from FAPERJ~(210.570/2024), CNPq~(305638/2023-8) and the CAPES-COFECUB~(20232475706P) program.

\paragraph{Data availability.} The data that support the findings of this Letter are available upon request. 

\bibliographystyle{apsrev4-2}
\bibliography{FilocheBib}

\end{document}